\documentclass[11pt,tightenlines,prd,aps,amsmath,amssymb,nofootinbib,floatfix,superscriptaddress,eqsecnum]{revtex4}

\usepackage{dcolumn}
\usepackage{graphicx}
\usepackage[mathscr]{euscript}
\usepackage[pdftex]{color}
\usepackage[sort&compress]{natbib}
\usepackage[colorlinks=true,linkcolor=blue,filecolor=blue,urlcolor=blue,citecolor=blue,pdftex,plainpages=false]{hyperref}
\usepackage{ulem}
\usepackage{slashbox}

\newcommand*{\ie}{\textit{i.e.},\ }

\newcommand{\cO}{\ensuremath{\mathcal{O}}}

\newcommand{\eq}[1]{Eq.~\eqref{eq:#1}}

\newcommand{\eqs}[2]{Eqs.~\eqref{eq:#1} and \eqref{eq:#2}}

\newcommand{\pole}{{\text{pole}}}
\newcommand{\MSbar}{{\overline{\text{MS}}}}
\newcommand{\mbar}{{\overline{m}}}

\newcommand{\be}{\begin{equation}}
\newcommand{\ee}{\end{equation}}
\newcommand{\bea}{\begin{eqnarray}}
\newcommand{\eea}{\end{eqnarray}}

\newcommand{\fourth}{\mbox{$\textstyle{\frac{1}{4}}$}}

\begin{document}

\title{A discussion on leading renormalon in the pole mass}
\author{J.~Komijani}
\email[]{j.komijani@tum.de}
\affiliation{Physik-Department, Technische Universit\"at M\"unchen, James-Franck-Stra{\ss}e 1, 85748 Garching, Germany}
\affiliation{Institute for Advanced Study, Technische Universit\"at M\"unchen, Lichtenbergstra{\ss}e 2a, 85748 Garching, Germany}

\date{\today}

\begin{abstract} 
Perturbative series of some quantities in quantum field theories, 
such as the pole mass of a quark, suffer from a kind of divergence called renormalon divergence. 
In this paper, the leading renormalon in the pole mass is investigated,
and a map is introduced to suppress this renormalon. 
The inverse of the map is then used to generate the leading renormalon 
and obtain an expression to calculate its overall normalization.  
Finally, the overall normalization of the leading renormalon of the pole mass is calculated 
for several values of quark flavors. 

\end{abstract} 

\maketitle

\section{Introduction} 
Perturbative calculations in quantum field theories may lead to divergent series. 
As an example one can consider the pole mass of a quark in a gauge theory such as QCD. 
After renormalization, which subtracts the UV divergences, 
the pole mass can be defined at each finite order in perturbation theory,  
but it cannot be defined at all orders \cite{Bigi:1994em,Beneke:1994sw}.\footnote{The   
pole mass is also IR finite at each order in perturbation theory \cite{Kronfeld:1998di}.}
This can be explained by the fact that the pole mass of a quark 
is not a physical quantity in a confining theory like QCD, 
thus the perturbative series for the pole mass need not converge. 
In addition to the pole mass, many perturbative calculations in physics yield divergent series 
even if the quantities of interest are finite and well-defined. 
For instance, the ground-state energy of the anharmonic oscillator
\bea 
  \Bigl(-\frac{d^2}{dx^2} + \fourth x^2 + \fourth \lambda\, x^4\Bigr) \Phi(x) 
  &=& E(\lambda)\, \Phi(x) 
\eea 
can be expanded in powers of the coupling constant $\lambda$, 
but it is not convergent for any $\lambda\neq0$~\cite{Bender:1969si}. 

There are some summation methods that can be used to handle divergent series. 
For instance, one can use the method of {\it Borel resummation} 
to assign analytic functions to a class of divergent series. 
(See Ref.~\cite{kawai2005algebraic} for a concise description of this method.) 
The Borel sum of a divergent series involves an integration in the Borel plane from a base point to infinity. 
The integration path is usually defined on the real axis from the origin to $+\infty$. 
But the choice of the integration path depends on the parameters of the series.  
In general, this method gives multiple Borel sums (for a divergent series) when 
there are multiple integration paths that cannot be deformed to each other. 
Particularly, this method may lead to ambiguous results when there are singularities 
on the positive real axis of the Borel plane. 
From physical point of view, the singularities in the Borel plane can have different origins.  
The factorial growth of number of Feynman diagrams at each order of perturbation theory
is a known source of these singularities. 
In theories with a running coupling constant, there is also a different set of singularities 
that stem from the very fact that the coupling constant runs. 
These singularities are called {\it renormalon} singularities,  
and occur in quantities such as the pole mass of a quark in QCD.\footnote{Ref.~\cite{Beneke:1998ui} 
defines the term {\it renormalon} as ``a singularity of the Borel transform related to 
large or small loop momentum behavior''. See also Ref.~\cite{Beneke:1998ui} for a discussion 
on the different sets of singularities in the Borel plane.}  

The pole mass can be expanded in powers of the coupling constant $\alpha$ as follows  
\bea\label{eq:m-pole:expansion}
  m_{\pole} = \mbar\left(1+\sum_{n=0}^{\infty} r_n {\alpha}^{n+1}(\mbar)\right)\, .  
\eea
Here  $\mbar$ denotes the $\MSbar$-renormalized mass at the scale $\mu=\mbar$. 
The coupling constant $\alpha$ is also in the same scheme and scale.  
It turns out that for large values of $n$ the coefficients $r_n$ grow roughly like $n!$, 
and thus the power expansion \eq{m-pole:expansion} is divergent. 
The large $n$ behavior of $r_n$ corresponds to a renormalon singularity in the Borel plane. 
It is known that the leading renormalon of this expansion is independent of the mass $\mbar$ \cite{Beneke:1994rs}. 
This statement immediately implies that the derivative of $m_{\pole}$ with respect to $\mbar$ 
is free of the leading renormalon. 
This observation can be used as a starting point to develop a method to investigate the leading renormalon of the pole 
mass. Such a method is the subject of this paper.

The derivative of \eq{m-pole:expansion} with respect to $\mbar$ reads  
\bea
  \frac{d m_{\pole}}{d \mbar} = 1 +  y + 2\beta(\alpha) y' \, ,
\eea
where 
\bea
  y &=& \sum_{n=0}^{\infty} r_n {\alpha}^{n+1}\, , \label{eq:y:expand:r_n} \\
  \beta(\alpha) &=& \frac{1}{2}\frac{d \alpha(\mbar)}{d \ln(\mbar)}
      = -\left(\beta_0 \alpha^2 + \beta_1 \alpha^3 + \beta_2 \alpha^4 + \cdots\right) \, . 
  \label{eq:beta:expansion}
\eea
One can calculate $y$ by solving the differential equation 
\be
  y + 2\beta(\alpha) y' = f(\alpha), \label{eq:ODE:y:f:1}
\ee
where
\be
  f(\alpha) \equiv \frac{d m_{\pole}}{d \mbar} - 1, \phantom{x} f(0)=0\, .
  \label{eq:define:f}
\ee
\eq{ODE:y:f:1} is a first order differential equation with the following solution
\be
  y(\alpha) = \int_{\alpha_\text{base}}^{\alpha} 
  \frac{d\alpha'}{2\beta(\alpha')}\, f(\alpha')\,
  \exp\left(-\int_{\alpha'}^{\alpha} \frac{d\alpha''}{2\beta(\alpha'')}\right)\, .
  \label{eq:y:f:1}
\ee
Note that $f(\alpha)$ is free of the leading renormalon of the pole mass; 
hence, this renormalon is generated through evaluating the integral in \eq{y:f:1}. 
Therefore, one can study the leading renormalon of the pole mass only 
by investigating the integral in \eq{y:f:1}, or equivalently  the differential equation in \eq{ODE:y:f:1}. 

I proceed in two different ways to investigate the leading renormalon of the pole mass. 
First, I develop a linear recurrence relation that reveals the structure of 
the leading renormalon as a function of the coefficients of the beta function. 
But, due to its linear nature, this recurrence relation leaves an undetermined overall normalization.  
Next, I solve \eq{ODE:y:f:1} and expand the solution in powers of $\alpha$ and then directly 
determine the large $n$ behavior of the expansion coefficients to obtain the overall normalization as well. 
Putting these two ways together, this paper presents a method to study the renormalon divergence 
of quantities such as the pole mass. 
It should be emphasized that the calculations are perturbative in nature. 

This paper is organized as follows. 
Section \ref{sec:theoretical_discussion} gives the details of the method.  
Section \ref{section:numerical_calculation} gives the overall normalization 
for several quark flavors and extends the discussion to the conformal window of QCD.  

\section{Renormalon in the pole mass: Theoretical Discussion}
\label{sec:theoretical_discussion}
\subsection{Leading renormalon from recurrence relation}
\label{subsec:recurrence}
It is known that the coefficients $r_n$,  in \eq{m-pole:expansion}, grow factorially as $n$ tends to infinity. 
In this subsection, I derive the large $n$ behavior of $r_n$, which is set by the coefficients of the beta function.  
First I calculate $f(\alpha)$, defined in \eq{define:f}, as follows
\bea
  f(\alpha) &=& \frac{d m_{\pole}}{d \mbar} - 1 \nonumber \\
        &=& y + 2\beta(\alpha) y'  \nonumber \\
        &=& \sum_{n=0}^{\infty} r_n {\alpha}^{n+1} - 2\left(\sum_{i=0}^{\infty} \beta_i 
  \alpha^{2+i}\right)\,\left(\sum_{n=0}^{\infty} r_n (n+1) {\alpha}^{n}\right) \nonumber \\            
        &=& \sum_{n=0}^{\infty} \Bigl(r_n - 2\bigl(\beta_0 n r_{n-1} + \beta_1 (n-1) r_{n-2} + \cdots + 
        \beta_{n-1} r_0 \bigr)\Bigr)\,{\alpha}^{n+1} \, ;
  \label{eq:expand:f:r_n}
\eea 
the power expansion of $f(\alpha)$ reads 
\be
  f(\alpha) = \sum_{k=0}^{\infty} r'_k\, \alpha^{k+1}\, ,
  \label{eq:define:f:expansion}
\ee
where
\bea
  r'_k  &=& r_k - 2\Bigl(\beta_0\, k\, r_{k-1} + \beta_1 (k-1) r_{k-2} + \cdots +  \beta_{k-1} r_0\Bigr)\, .
  \label{eq:r_n_prime:r_n}
\eea	      
Given the case that $f(\alpha)$ is free of the leading renormalon of $y(\alpha)$, 
the corresponding divergence in $r_n$ must be exactly canceled in the right side of \eq{r_n_prime:r_n}. 
This amount of information is enough to specify the pattern of divergence of $r_n$ for large values of $n$. 

The case that the leading renormalon divergence of $r_n$ for large values of $n$ 
cannot propagate to $r'_k$ through \eq{r_n_prime:r_n} can be expressed different. 
Let us focus on the quantity in parentheses in the coefficient of $\alpha^{n+1}$ in \eq{expand:f:r_n}.  
It implies a recurrence relation 
\be
  a_n = 2\Bigl(\beta_0\, n\, a_{n-1} + \beta_1 (n-1) a_{n-2} + \cdots +  \beta_{n-1} a_0\Bigr)\,,\phantom{xx} n\ge1\,,
  \label{eq:r_n:kernel}
\ee
which has a solution that diverges as $n\to\infty$, but such a solution cannot propagate 
to the sequence $r'_k$ through \eq{r_n_prime:r_n}.
Putting an overall constant aside, the recurrence relation in \eq{r_n:kernel} has only one solution. 
The large $n$ behavior of this solution can be determined using the ansatz  
\bea
  a_n &=& N\, (2\beta_0)^n\, \frac{\Gamma(n{+}1{+}b)}{\Gamma(1{+}b)} 
  \left(1 + \frac{s_1}{n{+}b} + \frac{s_2}{(n{+}b)(n{+}b{-}1)} + \frac{s_3}{(n{+}b)(n{+}b{-}1)(n{+}b{-}2)} 
  +  \cdots \right) . \nonumber \\
  \label{eq:ansatz:a_n}
\eea
Plugging this ansatz to \eq{r_n:kernel}, one then obtains 
\bea
  b   &=& \frac{\beta_1}{2\beta_0^2}\, , \\
  s_1 &=&  b^2 - c_2\, , \\
  s_2 &=& \bigl((b^2-c_2)^2 - b^3  + 2 b c_2  - c_3\bigr)/2\, , \\
  s_3 &=& \bigl((b^2 -c_2)^3 - 3(b^2-c_2)(b^3 - 2 b c_2 + c_3) + 2 b^4 - 6 b^2 c_2 + 2 c_2^2 + 4 b c_3
	  - 2 c_4\bigr)/6\,,
\eea
where $b$ is not a negative integer and 
\bea
  c_2 = \frac{\beta_2}{4\beta_0^3}\,,\phantom{x} c_3 = \frac{\beta_3}{8\beta_0^4}\,,\phantom{x}
  c_4 = \frac{\beta_4}{16\beta_0^5}\, .
\eea
This result is identical to the leading renormalon in the pole mass
(see for instance Refs.~\cite{Beneke:1994rs,Beneke:1998ui,Ayala:2014yxa}). 
It is straightforward to improve the ansatz by including more terms in \eq{ansatz:a_n}
and calculating their corresponding coefficients. 

The first two coefficients in the beta function, namely $\beta_0$ and $\beta_1$, are independent of the choice of 
the scheme. The higher order coefficients, which are scheme dependent, are usually given in the $\MSbar$ scheme. 
But, in order to simplify the calculations, one can employ schemes in which the higher order coefficients have 
simple forms. Below, two special schemes are discussed. 
First, consider the scheme in which 
\be
\beta(\alpha) = - (\beta_0 \alpha^2 + \beta_1 \alpha^3) \, .
\label{eq:beta:scheme:0:1}
\ee
To solve \eq{r_n:kernel} with this scheme, I use the $z$-transform\footnote{See 
Ref.~\cite{oppenheim2010discrete} for the definition and applications of the $z$-transform.} 
and after a change of variable as $u=1/z$, I obtain 
\be
 \label{eq:a_n:beta-scheme:0:1:solution} 
 a_n =   (2\beta_0)^n\,  n!\, a_0\,
 \oint_C \frac{du}{2\pi i}\, \frac{1}{u^{n+1}}\, \frac{e^{-u b}}{(1-u)^{1+b}} \, ,
\ee
where $C$ is a counter-clockwise closed path encircling the origin 
and crossing the right side of the real axis at $u_c \in (0,1)$. 
For a non-integer value of $b$, the branch cut of the integrand is assumed to be 
on the real axis from $u=1$ to $+\infty$. 
One can easily verify that \eq{a_n:beta-scheme:0:1:solution} yields the ansatz 
given in \eq{ansatz:a_n} with $\beta_n=0$ for $n>1$. 
When $b$ is a negative integer, the factorial divergence disappears.
As second special scheme, let us consider a scheme in which the beta function has the form 
\be
  \beta(\alpha) = \frac{- \beta_0 \alpha^2}{1 - (\beta_1/\beta_0) \alpha} \, .
  \label{eq:beta:scheme:0:1-inv}
\ee 
This choice of the beta function, which will prove convenient in reducing the algebra,  
has been employed in the literature in studies of renormalons. (See for instance Refs.~\cite{Brown:1992pk, Lee:1996yk}.)  
For this scheme, the exact solution of the recurrence relation in \eq{r_n:kernel} is
\be
 a_n = (2\beta_0)^n\, \frac{\Gamma(n+1+b)}{\Gamma(2+b)}\, a_0 \, , \phantom{xx} n\ge1\, .
 \label{eq:a_n:beta-scheme:0:1-inv:solution}
\ee
Note that, by setting $\beta_n=\beta_0 (\beta_1/\beta_0)^n$ for $n>1$,  
the ansatz given in \eq{ansatz:a_n} reduces to \eq{a_n:beta-scheme:0:1-inv:solution}. 
One can also start from \eq{a_n:beta-scheme:0:1-inv:solution} and, after a scheme conversion, 
derive the ansatz in \eq{ansatz:a_n}.
 
I use the expression ``pure-renormalon sequence'' to refer to the sequence $a_n$, 
which is the solution of the recurrence relation in \eq{r_n:kernel}. 
The fact that the pure-renormalon sequence (up to a constant) only depends on 
the coefficients of the beta function is not surprising because renormalons are related to the notion of 
the running coupling constant~\cite{Beneke:1998ui} and the running is governed by the beta function. 
For the sake of simplicity, hereafter the discussion is restricted to the scheme 
with the beta function given in \eq{beta:scheme:0:1-inv} unless otherwise stated. 
Having the sequence $a_n$ determined in \eq{a_n:beta-scheme:0:1-inv:solution}, 
the objective is now to calculate the overall normalization $a_0$ such that
\be
  r_n \sim a_n\, ,
\ee
as $n\to\infty$. The conventional overall normalization $N$, which is more often used in the literature, is then 
\be 
  N \equiv \frac{a_0}{1+b}\label{eq:a0:2:N}\, .
\ee
Note that the overall normalization $N$ is not necessarily invariant under a change in the scheme and scale. 
The scheme conversion is discussed below.

\subsection{Overall normalization of the leading renormalon}
\label{subsec:overall-normalization} 
It was discussed that the renormalon divergence is produced in the process of 
evaluating the integral in \eq{y:f:1}, and replacing it with a power series. 
Instead of working with \eq{y:f:1}, which gives the integral representation of the solution of \eq{ODE:y:f:1}, 
it is easier to use the formal solution  
\bea
 y(\alpha) &=& \frac{1}{1 + 2\beta(\alpha)\frac{d}{d\alpha}} f(\alpha)\, .
 \label{eq:IE:y:exact:3}
\eea
As mentioned above the discussion is restricted to a scheme in which the beta function 
is given in \eq{beta:scheme:0:1-inv}.
To this end, one can use the expansion 
\be
  \alpha_{\MSbar} = \alpha
    + \Bigl(\frac{\beta_2}{\beta_0} - \bigl(\frac{\beta_1}{\beta_0}\bigr)^2\Bigr)\alpha^3 
    + \frac{1}{2}\Bigl(\frac{\beta_3}{\beta_0} - \bigl(\frac{\beta_1}{\beta_0}\bigr)^3\Bigr)\alpha^4 + \cdots 
    \label{eq:alpha-MSbar:2:alpha}
\ee
to convert a series in powers of the coupling constant $\alpha_\MSbar$ to a series in powers of $\alpha$.
Defining $ u = \beta_0/(\beta_1\alpha)$, \eq{IE:y:exact:3} reads
\bea
y(\alpha) &=&  \frac{1}{1 + \frac{b^{-1}}{1-u^{-1}}\frac{d}{du} }\, f(\alpha)
	   \nonumber \\
	   &=&  \frac{1}{1 + \frac{b^{-1}}{1-u^{-1}}\frac{d}{du} }\, 
	   \sum_{k=0}^\infty r'_k\, \Bigl(\frac{\beta_1}{\beta_0}\,u\Bigr)^{-(1+k)}\, .
 \label{eq:IE:y:exact:4}
\eea
Simplifying this formal expression, one can calculate the large-$n$ behavior of the sequence $r_n$. 
This is discussed in detail in the Appendix. 
Exploiting \eq{app:a_n:asymptotic:3}, in the Appendix,  and setting
\be
    \nu = 1\,,\quad 
    x   = b\,,\quad 
    z   = u\,,\quad   
    d_k = r'_k\, \bigl({\beta_0}/{\beta_1}\bigr)^{1+k}\,,\quad 
    a_n = r_n\,  \bigl({\beta_0}/{\beta_1}\bigr)^{1+n}\,,
\ee
$r_n$ reads 
\bea
  r_n &\sim& \Gamma(1 + n + b)\,b^{-n}\,({\beta_1}/{\beta_0})^{1+n}\, \sum_{k=0}^\infty r'_k\,  
	      \frac{(1+k)\, b^k}{\Gamma(2+k+b)}\, \bigl({\beta_0}/{\beta_1}\bigr)^{1+k}\nonumber \\
   &\sim& \frac{\Gamma(1 + n + b)}{\Gamma(2+b)} \bigl(2\beta_0\bigr)^{n} 
   \sum_{k=0}^\infty r'_k\,  \frac{(1+k)\, \Gamma(2+b)}{\Gamma(2+k+b)}\, \bigl(2\beta_0\bigr)^{-k}  
    \phantom{xxxxx}\,(n\to\infty)\, .
  \label{eq:app:r_n:asymptotic:final}
\eea
Recalling \eqs{a_n:beta-scheme:0:1-inv:solution}{a0:2:N}, 
the overall normalization of the leading renormalon of the pole mass is then 
\bea
  \label{eq:renormalon_constant}
  N  &=&  \frac{a_0}{1+b} = 
  \sum_{k=0}^\infty r'_k\,  \frac{\Gamma(1+b)}{\Gamma(2+k+b)}\, \frac{1+k}{(2\beta_0)^{k}}\,\, . 
\eea
Note that because $r'_k$ are free of the leading renormalon in the pole mass, 
the large $k$ behavior of $r'_k$ is governed by higher order renormalons in the pole mass,
which grow roughly as $(\beta_0)^k ~\Gamma(1+k)$.\footnote{Here
I assume that the leading and next-to-leading renormalons are the dominant sources of divergence. 
See Ref.~\cite{Beneke:1998ui} for the discussion on the different sets of known singularities
in the Borel plane and their distance from the origin of the Borel plane.} 
Therefore the expression in \eq{renormalon_constant} converges. 
In practice, one needs to truncate the series and calculate $N$ using
\bea
  \label{eq:renormalon_constant_truncated}
  N_{k_\text{max}}  &=&  \sum_{k=0}^{k_\text{max}} r'_k\,  \frac{\Gamma(1+b)}{\Gamma(2+k+b)}\, 
  \frac{1+k}{(2\beta_0)^{k}}\,\, . 
\eea

The constant $N$ is not independent of the scheme. 
But, as discussed in \cite{Beneke:1998ui}, $N$ is invariant if the coupling constant of 
two schemes, denoted by $\alpha$ and $\tilde\alpha$, 
are related by $\alpha = \tilde\alpha + \cO(\tilde\alpha^3)$\ . 
Note that there are several works that use different methods and present different series 
to calculate the overall normalization of the leading renormalon~\cite{Lee:1999ws, Pineda:2001zq, Hoang:2008yj}. 
A quick comparison shows that their truncated series are not identical with \eq{renormalon_constant_truncated}. 

\section{Renormalon in the pole mass: Numerical Calculations}
\label{section:numerical_calculation}
\subsection{Large number of flavor}
\label{subsection:large-flavor}
Now I investigate \eq{renormalon_constant} in the limit of large number of flavors. 
At leading order in this limit, one can keep only $\beta_0$ and drop all $\beta_n$ for $n>0$ and set $b=0$. 
Then, \eq{renormalon_constant} reads
\bea
  \label{eq:renormalon_constant:large:flavor}
   \Bigl. N \Bigr|_{(\text{large } n_f)}  
    &=&  \sum_{k=0}^\infty  r_k^\prime\, \frac{1}{k!}\, \frac{1}{(2\beta_0)^{k}}\nonumber \\
    &=&  r_0 + \sum_{k=1}^\infty \bigl(r_k - 2\beta_0\,k\,r_{k-1}\bigr)\, \frac{1}{k!}\, 
\frac{1}{(2\beta_0)^{k}}\nonumber \\
     &=&  \left. \Bigl( (1-2u)\,B[y](u)\Bigr) \right|_{u=1/2} \nonumber \\ 
     &=&  \frac{4}{3\pi}\,e^{5/6} \, ,
\eea
where $B[y](u)$ is the Borel transform of $y = \sum_{n=0}^{\infty} r_n {\alpha}^{n+1}$, which is 
\bea 
 B[y](u) &=& \sum_{n=0}^{\infty} \frac{r_n}{n!}\, \bigl(\frac{u}{\beta_0}\bigr)^{n}\, .
\eea
Note that this is identical to $B[\delta m/m](u)$ defined in Eq.~(4.3) of Ref.~\cite{Ball:1995ni}, \ie
\bea
 B[y](u) = B[\delta m/m](u) 
  = \frac{1}{3\pi}\left(6\, e^{5u/3} (1-u)\frac{\Gamma(u)\Gamma(1-2u)}{\Gamma(3-u)} + \frac{\tilde G_0(u)}{u}\right)\,,
\eea
where $\tilde G_0(u)$ is finite at $u=1/2$. 

One might wonder how \eq{renormalon_constant} numerically converges to $N$ for large number of flavors.
Using the numerical values of $r_n$ provided in Table 1 of Ref.~\cite{Beneke:1994qe}, 
\eq{renormalon_constant_truncated} gives 
\be
  [0.4244,\, 0.9944,\, 0.9349,\, 0.9714,\, 0.9659,\, 0.9770,\, 0.9746,\, 0.9769,\, 0.9762]\, , 
  \label{eq:large:flavor:successive}
\ee
for $k_{\rm max}=0,1,\cdots,8$. 
Note that the last number, $N\approx0.9762$, is close to the exact result
\be 
  \frac{4}{3\pi}\,e^{5/6}\, \approx\, 0.97656\, .
\ee

\subsection{Finite number of flavors}

The relation between the pole mass and the $\MSbar$ mass is known up to order $\alpha_s^4$~\cite{Marquard:2016dcn}.  
In this subsection, I use this relation to calculate $r'_k$  for several values of $n_l$ ranging from 0 to 6, 
and determine the overall normalization of the leading renormalon using the truncated expression in 
\eq{renormalon_constant_truncated}. 
The results, for $k_\text{max}$ from 0 to 3, are listed in Table \ref{table:N:nl}. 
For each number of flavors, I take the last column of Table \ref{table:N:nl} as the central value of $N$ 
and twice of the difference of the last two columns as a conservative estimate of the truncation error. 
For instance, for $n_l=3$, we obtain   
\begin{equation}
 N = 0.535 \pm 0.010 \, . 
\end{equation} 
There are several calculations of the overall normalization of the leading renormalon in the pole mass. 
For some recent calculation see Refs.~\cite{Ayala:2014yxa, Beneke:2016cbu}. 
Considering the uncertainties, the results of this paper are in agreement with them. 

\begin{table}[t]
\begin{tabular}{|c||c|c|c|c|}
\hline 
\backslashbox{\,\,\,\,$n_l$}{$k_\text{max}$}
  & \phantom{xx}  0 \phantom{xx} &   \phantom{xx}  1 \phantom{xx}
  & \phantom{xx}  2 \phantom{xx} &   \phantom{xx}  3 \phantom{xx}  \\ 
\hline
\hline
 0  &  0.299   &  0.501   &  0.577  & 0.592 \\ 
\hline 
 1  &  0.299   &  0.494   &  0.566  & 0.576 \\ 
\hline 
 2  &  0.301   &  0.487   &  0.554  & 0.558 \\ 
\hline 
 3  &  0.304   &  0.483   &  0.539  & 0.535 \\ 
\hline 
 4  &  0.310   &  0.480   &  0.522  & 0.505 \\ 
\hline 
 5  &  0.319   &  0.482   &  0.498  & 0.463 \\ 
\hline 
 6  &  0.335   &  0.489   &  0.461  & 0.396 \\ 
\hline 
\end{tabular}
  \caption{
    The values of $N$ obtained from the truncated expression in \eq{renormalon_constant_truncated} 
    for $k_\text{max}$ from 0 to 3, and for $n_l$ from 0 to 6.
  }
  \label{table:N:nl}
\end{table}

\subsection{Near conformal window of QCD}

Now I discuss the leading renormalon in a region close to the conformal window of QCD. 
The first two coefficients of the beta function, namely $\beta_0$ and $\beta_1$, are scheme independent 
and they are positive for small values of flavors. 
There is a region in which $\beta_0$ is positive and $\beta_1$ is negative,  
which indicates the presence of a non-trivial zero in the beta function in this region \cite{Banks:1981nn}. 
One can use \eq{renormalon_constant} to study the leading renormalon in this region, 
but this relation should be treated carefully for $n_l$ at vicinity of $16.5$, where $\beta_0$ vanishes and $b$ 
blows up. In other words, the assumptions under which 
\eq{renormalon_constant} is derived might be problematic when $\beta_0$ vanishes and $b$ becomes large. 
Here I discuss two important possible obstacles in calculations of $N$ in this region. 
First, the factorial growth of the coefficients due to the leading renormalon 
appears only for large values of $n$. A lower limit of $n$ for which the factorial growth is noticeable depends on $b$. 
A rough estimate for the lower limit of $n$ can be obtained based on a discussion in the Appendix. 
Taking advantage of \eq{app:condition:n:x}, the heuristic condition for $n$ is 
\be
  |b| < 1+n\, . 
  \label{eq:condition:b:n} 
\ee
As this condition implies, when $b$ becomes large, 
the pattern of factorial growth in the coefficients appears only for very large values of $n$. 
Therefore, any method that estimates $N$ by comparing the exactly known coefficients $r_n$ 
and the expectations based on the leading renormalon may fail if $n$ is not large enough. 
For instance, Ref. \cite{Ayala:2014yxa} uses such a comparison 
and finds that $N$ tends to zero in the range $n_l\in(12,23)$. 
Ref. \cite{Beneke:2016cbu} confirms this behavior, 
however it is then discussed that the extracted value of $N$ is completely unreliable in this region, 
and the smallness of $N$ is therefore a technical artifact of the method they use, 
which ceases to be valid when $b$ becomes large. 
This argument is consistent with the spirit of \eq{condition:b:n}. 

The other important thing to be discussed is that \eq{renormalon_constant} is basically derived for 
the scheme with beta function defined in \eq{beta:scheme:0:1-inv}, \ie $r'_k$ are the coefficients
of the expansion in powers of $\alpha$ with the beta function given in \eq{beta:scheme:0:1-inv}. 
As discussed before, one should use \eq{alpha-MSbar:2:alpha} 
to convert a series in powers of $\alpha_\MSbar$ to a series in powers of $\alpha$. 
As it is evident from \eq{alpha-MSbar:2:alpha}, this conversion should be treated carefully 
when $\beta_0$ is nearly zero. Indeed, the conversion of the schemes becomes singular when $\beta_0$ vanishes 
and consequently the scheme dependence of $N$ should be treated carefully.  

Before discussing the scheme conversion let us write \eq{renormalon_constant} in a form 
useful for the current discussion. 
For large values of $b$, one can follow the discussion in the Appendix and take advantage of 
\eq{app:a_n:asymptotic:4} to show that 
\eq{renormalon_constant} can be expanded as 
\bea
  \label{eq:renormalon_constant:large_b:f-prime}
  N  &=& \left.
    \left(\frac{1}{1 + b}f'(\alpha) + \cdots \right)\right|_{\alpha=1/b_1} \, .
\eea
Recall that the coupling constant $\alpha$ is supposed to be 
in the scheme with the beta function given in \eq{beta:scheme:0:1-inv}. 
In order to proceed, it must be discussed carefully how to obtain $f(\alpha)$ from $f_\MSbar(\alpha_\MSbar)$, 
which is the corresponding expression in the $\MSbar$ scheme. 

For the sake of simplicity, let us define an intermediate scheme in which the beta function is given 
in \eq{beta:scheme:0:1} and the coupling constant is denoted by $\overline\alpha$, 
and convert the expression for the pole mass that is given as a series in powers of $\alpha_\MSbar$ 
to a series in powers of $\overline\alpha$ using
\be
    \alpha_{\MSbar} = \overline\alpha
    + \frac{\beta_2}{\beta_0}\, \overline\alpha^3 
    + \frac{1}{2}\frac{\beta_3}{\beta_0}\, \overline\alpha^4 + \cdots\, . 
    \label{eq:alpha-MSbar:2:alpha-bar}
\ee
Note that this relation does not make any problem in the calculation of $N$ even for very small values of $\beta_0$.
The main difficulty appears in expressing $\overline\alpha$ in terms of $\alpha$. 
Let us define the auxiliary independent and dependent variables $t$ and $T$ as follows
\bea
  t &\equiv& b_1 \alpha \\
  T(t) &\equiv& b_1 \overline\alpha = \sum_{n=0}^\infty c_n t^n \,, \label{eq:T:Taylor}
\eea
where $c_0=1$ and $c_1=0$. Using the beta functions of both schemes, one can easily 
derive a differential equation to calculate $T(t)$ as follows
\bea
  \frac{d\overline\alpha}{d \ln(\mu)} &=& \frac{d\alpha}{d \ln(\mu)} \frac{dT}{dt}\\
  \Rightarrow\quad T^2(1+T) &=& \frac{t^2}{1-t}\, T' \, .
\eea
The family of solutions that are analytic at the origin have expansions as $T(t) = t + \cO(t^2)$. 
Solving the differential equation and imposing the condition $c_1=0$, 
one obtains  
\bea
  \bigl(1+\frac{1}{T}\bigr)\,e^{-\frac{1}{T}} &=& \frac{1}{t}\,e^{-\frac{1}{t}}\, .
\eea
The solution to this equation can be written in terms of the Lambert $W$ function as\footnote{See 
Ref.~\cite{corless1996lambertw} for the definition and properties of the Lambert W function.}
\bea 
  T(t) &=& \frac{-1}{1+ W(-\frac{1}{t}e^{-\frac{1}{t}-1})}\, . 
  \label{eq:T:by:W}
\eea 
In order to avoid the discussion on different Riemann sheets of the $W$ function, 
I use an alternative representation
\bea
T(t) \equiv \frac{t}{1 + t X(t)}\, , \label{eq:T:2:X}
\eea 
where $X(t)$ is implicitly defined by
\bea 
  e^X = 1 + t + t X \, .
\eea 
Here, only the solution of $X(t)$ that is analytic at the origin is of interest. 
The Taylor expansion of this solution reads
\bea 
  X(t) = t + \frac{1}{2}t^2 + \cdots \, \label{eq:X:Taylor} .
\eea 
Both $X(t)$ and $T(t)$ are singular at $t = -1/W(-1)$, 
which restricts the convergence of the Taylor expansions in \eqs{T:Taylor}{X:Taylor} to 
\be 
  |t| \lesssim 0.7275  \, .
\ee
Therefore, the Taylor expansion in \eq{T:Taylor} diverges at $t=1$, \ie $\alpha=1/b_1$. 
This is the very point in which \eq{renormalon_constant:large_b:f-prime} should be evaluated. 
Thus, the sequence of scheme conversions 
\be
  f_\MSbar(\alpha_\MSbar)\quad \leftrightarrow\quad
  \overline f(\overline\alpha)\quad \leftrightarrow\quad
  f(\alpha)\,\, ,
\ee
is problematic at $\alpha=1/b_1$ if one wishes to use Taylor expansions for the scheme conversions. 

\section{Conclusion}
In this paper, I introduced a method to study the leading renormalon in the pole mass. 
This method yields a linear recurrence relation that reveals the structure of the leading renormalon. 
The recurrence relation depends only on the coefficients of the beta function. 
This is not surprising because renormalons are related to the notion of 
running coupling constants and the running is governed by the beta function.
This method also gives an expression to calculate the overall normalization of the leading renormalon. 
The overall normalization of the leading renormalon of the pole mass was then 
calculated for several values of quark flavors, and was discussed for the near-conformal window of QCD. 

\section{Acknowledgments}  
I thank Martin Beneke for helpful discussions and comments on earlier version of the manuscript.   
I am also grateful to Nora Brambilla, Andreas S. Kronfeld and Antonio Vairo for helpful conversations 
and comments on the manuscript. 
This work is supported in part by the German Excellence Initiative and the European Union Seventh Framework
Programme under grant agreement No.~291763.

\begin{appendix}

\section{Large-order behavior of an asymptotic expansion}
\label{Appendix:1}

In this appendix we investigate the expression  
\be
  F(z,x) = \frac{1}{1+\frac{x^{-1}}{1-z^{-1}}\,\frac{d}{dz}}\,  f_0(z) \, , 
  \label{eq:app:F:formal}
\ee
which is the formal solution of the first order differential equation 
\be
  \left(1+\frac{x^{-1}}{1-z^{-1}}\,\frac{d}{dz}\right) F(z,x) = f_0(z) \, .
  \label{eq:app:F:ode}
\ee
The solution can be expanded as 
\be
  F(z,x) = \sum_{n=0}^{\infty} f_n(z)\,  x^{-n} \, ,
  \label{eq:app:F:def}
\ee
where
\bea
  f_n(z) &=& \Bigl(\frac{-1}{1-z^{-1}}\,\frac{d}{dz}\Bigr)^n\, f_0(z)\, .
  \label{eq:app:f:def}
\eea
Given $f_0(z)$, one can derive $f_n(z)$ and in turn calculate $F(z,x)$. 

This problem can be tackled using integral representation of $F(z,x)$, which involves the Lambert $W$ function.
But in this appendix we wish to work explicitly with \eqs{app:F:def}{app:f:def}. 
First, let us introduce a set of formal series defined in terms of the gamma function and its derivatives as
\bea
  g_n(z; \nu) &\equiv& \sum_{k=0}^{\infty} \frac{\Gamma^{(k)}(\nu+n+k)}{\Gamma(\nu)\Gamma(k+1)}\, z^{-(\nu+n+k)}  \, .
  \label{eq:app:g}
\eea
These set of series can be generated from $g_0(z; \nu)$, using the following relation 
\bea
  g_{n}(z; \nu) &=& \Bigl(\frac{-1}{1-z^{-1}}\, \frac{d}{dz}\Bigr)^n\, g_{0}(z; \nu) \, .
\eea
This can be proved by induction as follows 
\bea
  \frac{-1}{1-z^{-1}}\, \frac{d}{dz}\, g_{n}(z; \nu) &=& \left(\sum_{j=0}^{\infty} z^{-j}\right) 
  \sum_{k=0}^{\infty} \frac{\Gamma^{(k)}(\nu+n+k)}{\Gamma(\nu)\Gamma(k+1)}\,(\nu+n+k)\, z^{-(\nu+n+1+k)}  \nonumber\\
    &=& \sum_{j=0}^{\infty} 
  \sum_{k=0}^{\infty} \frac{\Gamma^{(k)}(\nu+n+k)}{\Gamma(\nu)\Gamma(k+1)}\,(\nu+n+k)\, z^{-(\nu+n+1+k+j)}\nonumber\\
    &=& \sum_{m=0}^{\infty} \sum_{j=0}^{m} 
    \frac{\Gamma^{(m-j)}(\nu+n+m-j)}{\Gamma(\nu)\Gamma(m-j+1)}\,(\nu+n+m-j)\, z^{-(\nu+n+1+m)}\nonumber \\
   &=& \sum_{m=0}^{\infty} \frac{\Gamma^{(m)}(\nu+n+1+m)}{\Gamma(\nu)\Gamma(m+1)}\, z^{-(\nu+n+1+m)}  \nonumber\\
   &=& g_{n+1}(z;\nu)\,\, .
  \label{eq:app:g:n+1}
\eea
In the third equality we reordered the terms and defined $m=k+j$, 
and in the fourth equality we exploit the identity
\bea
 \Gamma^{(m)}(t + 1) &=& 
 \sum_{j=0}^{m} \, \frac{\Gamma(m+1)}{\Gamma(m+1-j)}\, (t-j)\, \Gamma^{(m-j)}(t-j) \, . 
 \label{eq:app:identity:I}
\eea
This identity can be proved as follows. 
Starting from $\Gamma(t+1)=t\,\Gamma(t)$, one can show
\bea
 \Gamma^{(m)}(t + 1 ) &=&  t\, \Gamma^{(m)}(t) + m\, \Gamma^{(m-1)}(t)\, .
 \label{eq:app:identity:I:proof:1}
\eea
Shifting the parameters and variables, one finds
\bea
 \Gamma^{(m-1)}(t) &=&  (t-1)\, \Gamma^{(m-1)}(t-1) + (m-1)\, \Gamma^{(m-2)}(t-1)\, .
\eea
Plugging this into \eq{app:identity:I:proof:1} yields
\bea
 \Gamma^{(m)}(t + 1 ) &=&  t\, \Gamma^{(m)}(t) + m\, (t-1)\,\Gamma^{(m-1)}(t-1) + m(m-1)\Gamma^{(m-2)}(t-1) \, .
\eea
It is straightforward to repeat the procedure and derive \eq{app:identity:I}.
Note that, using the integral representation of the gamma function, the series in \eq{app:g} can be summed up, 
which yields
\bea
  g_n(z; \nu) &=& \frac{z^{-(\nu+n)}}{\Gamma(\nu)}\, \int_0^\infty dt\, e^{-t + \frac{t}{z}\ln(t)} t^{n+\nu-1}  \, .
  \label{eq:app:g:integral}
\eea
Assuming $\Re(\nu+n)>0$, 
for any non-zero value of $z$, one can always choose the integration path to infinity 
in such a way that the integral remains finite. 
\eq{app:g:n+1} can be also verified using \eq{app:g:integral}. 

Now we return to \eqs{app:F:def}{app:f:def}, and calculated $F(z,x)$ for the case $f_0(z) = g_0(z;\nu)$.
For this special case we find that $f_n(z) = g_n(z;\nu)$ and consequently
\bea
F(z,x) &=& \sum_{n=0}^{\infty}  g_n(z;\nu)\,  x^{-n} \nonumber \\
      &=& \sum_{n=0}^{\infty} \sum_{k=0}^{\infty} 
	    \frac{\Gamma^{(k)}(\nu + n + k )}{\Gamma(\nu) \Gamma(k+1)} \, x^{-n}\, z^{-(\nu+n+k)} \nonumber \\
      &=& \sum_{m=0}^{\infty} \sum_{k=0}^{m}
	    \frac{\Gamma^{(k)}(\nu + m )}{\Gamma(\nu) \Gamma(k+1)} \, x^{-(m-k)}\, z^{-(\nu+m)} \nonumber \\
      &=& \sum_{m=0}^{\infty} \Bigl( 
	    \Gamma(\nu + m + x) - R_m(x;\nu+m) \,  \Bigr) \, 
	  \frac{x^{-m}}{\Gamma(\nu)}\, z^{-(\nu+m)} \, .
      \label{eq:app:F:calc}
\eea
In the third equality we reordered the terms and defined $m=n+k$, and in the fourth equality we defined
\bea	
    R_m(x;\nu+m) &\equiv& \int_0^x dt\, \frac{(x-t)^{m}}{\Gamma(m+1)} \Gamma^{(m+1)}(\nu + m + t)\, .
    \label{eq:app:R:Gamma}
\eea
Now constructing a power expansion for $F(z,x)$ as   
\bea
  F(z,x) &=& \sum_{n=0}^{\infty} a_n(x)\, z^{-(\nu+n)} \, ,
  \label{eq:app:F:expand:a_n}
\eea
we conclude that the large $n$ behavior of $a_n(x)$ is
\bea
  a_n(x) &\sim& \frac{\Gamma(\nu + n + x)}{\Gamma(\nu)} \, x^{-n}\, . 
  \label{eq:app:a_n:asymptotic}
\eea
Here $n$ is assumed to be large enough such that $R_n(x;\nu+n)$ is negligible compared to $\Gamma(\nu + n + x)$. 
This assumption is particularly very important when $x$ is a large number. 
Based on the convergence radius of the Taylor expansion of $\Gamma(y)$ about $y=\nu+n$,  
one can argue that $n$ must obey the condition 
\be
  |x| < \nu+n 
  \label{eq:app:condition:n:x}
\ee 
when $\nu$ is a real positive number. This condition restricts 
the range of $n$ that $a_n(x)$ behaves asymptotically as \eq{app:a_n:asymptotic}. 

Now we extend the calculation to the case that $f_0(z)$ is a linear combination of $g_0(z;\nu)$ as  
\bea
  f_0(z) &=& \sum_{l=0}^\infty c_l\, g_0(z;\nu+l)\, . 
  \label{eq:app:f:expand:g_k}
\eea
We wish to calculate $F(z,x)$, expand it as \eq{app:F:expand:a_n} and derive the large $n$ behavior of $a_n(x)$. 
Since we are dealing with a linear problem, it is straightforward to repeat the 
calculations and derive the large $n$ behavior of $a_n(x)$. We obtain 
\bea
  a_n(x) &\sim& \, \Gamma(\nu + n + x)\,x^{-n}\sum_{l=0}^{n}\, c_l\, \frac{ x^l}{\Gamma(\nu+l)} 
  \label{eq:app:a_n:asymptotic:2}
\eea 
as $n\to\infty$. This expression is derived based on the approximation that led to \eq{app:a_n:asymptotic}. 
This approximation might be problematic when $l\approx n$. 
However, we assume that the series in the above expression is finite when $n\to\infty$, 
and we assume that $n$ is large enough such that we can ignore the approximation. 
This issue will be addressed again in a discussion after \eq{app:a_n:asymptotic:3}. 

For any function that can be expanded as \eq{app:f:expand:g_k}, the large $n$ behavior of $a_n(x)$ can be calculated 
using \eq{app:a_n:asymptotic:2}. 
For instance, let us consider  
\bea
  f_0(z) &=& z^{-(\nu+m)}\, ,
\eea
where $m$ is a non-negative integer. Its expansion in terms of $g_0(z;\nu)$ reads
\bea 
   z^{-(\nu+m)} &=& \sum_{l=m}^\infty c_{(\nu,m,l)}\, g_0(z;\nu+l)\nonumber \\
	&=& \sum_{l=m}^\infty c_{(\nu,m,l)}\,
	\sum_{k=0}^{\infty} \frac{\Gamma^{(k)}(\nu+l+k)}{\Gamma(\nu+l)\Gamma(k+1)}\, z^{-(\nu+l+k)} \nonumber \\
	&=& \sum_{n=m}^\infty
	\sum_{k=0}^{n-m}  c_{(\nu,m,n-k)}\,\frac{\Gamma^{(k)}(\nu+n)}{\Gamma(\nu+n-k)\Gamma(k+1)}\, z^{-(\nu+n)} \, ,
\eea
where the unknown coefficients $c_{(\nu,m,l)}$ are defined for $l\ge m$, and they can be calculated from the 
linear algebraic system of equations 
\bea
  \sum_{k=0}^{n-m} c_{(\nu,m,n-k)} \frac{\Gamma^{(k)}(\nu+n)}{\Gamma(\nu+n-k)\Gamma(k+1)}  &=&  \delta_{n m}\, .
\eea
Considering the identity
\bea
\frac{y}{N!}\,\frac{d^N}{dy^N}\, \frac{\Gamma(y+N)}{\Gamma(y+1)} &=& \delta_{N 0}\,\, ,
\eea
and setting $N=n-m$, one finds that 
\bea
c_{(\nu,m,l)} &=& \frac{(\nu+m)\, \Gamma(\nu+l)}{\Gamma(l-m+1)}\, 
	\left.\left(\frac{d}{dy}\right)^{l-m}\frac{1}{\Gamma(y+1)}\right|_{y=\nu+m} \, ,
\eea
where $l\ge m$. We also define $c_{(\nu, m, l)}=0$ for $l<m$. 
Now one can immediately calculate the large $n$ behavior of $a_n(x)$ 
using \eq{app:a_n:asymptotic:2}. Thus, for $f_0(z) =  z^{-(\nu+m)}$, the result reads
\bea
  a_n(x) &\sim& \, \Gamma(\nu+n+x)\,x^{-n}\sum_{l=0}^{n} c_{(\nu,m,l)}\,\frac{x^l}{\Gamma(\nu+l)} \nonumber \\
	 &\sim& \, \Gamma(\nu+n+x)\,x^{-n+m}\,(\nu+m)\, \sum_{k=0}^{n-m} \frac{x^k}{\Gamma(k+1)}  
	\left.\left(\frac{d}{dy}\right)^{k}\frac{1}{\Gamma(y+1)}\right|_{y=\nu+m} \nonumber \\
	 &\sim& \, \frac{\Gamma(\nu+n+x)}{\Gamma(1+\nu+m+x)}\,(\nu+m)\, x^{-n+m} \phantom{xxxxx}\,(n\to\infty)\, .
  \label{eq:app:a_n:asymptotic:2.5}
\eea
To obtain the third line from the second line, we assumed that $(n-m)$ is large enough 
such that we can neglect the higher order terms in the Taylor expansion of the reciprocal gamma function. 
Note that the reciprocal gamma function is analytic at all finite points of the complex plane, 
therefore we cannot obtain an estimate for the lower value of $n$ similar to \eq{app:condition:n:x}. 

We now derive the large $n$ behavior of $a_n(x)$ for 
\bea
  f_0(z) &=& \sum_{m=0}^\infty\, d_m\, z^{-(\nu+m)}\, . 
  \label{eq:app:f:expand:c_m}
\eea
For this general case, \eq{app:a_n:asymptotic:2.5} implies 
\bea
  a_n(x) &\sim& \, \Gamma(\nu+n+x)\,x^{-n}\,\sum_{m=0}^{\infty}\, d_m\, \frac{(\nu+m)\,x^m}{\Gamma(1+\nu+m+x)}\, 
    \phantom{xxxxx}\,(n\to\infty)\, .
  \label{eq:app:a_n:asymptotic:3}
\eea
Note that this expression is derived based on the approximation that was used in derivation of 
\eq{app:a_n:asymptotic:2.5}, but that approximation is not correct when $m\gtrsim n$. 
However, we do not modify \eq{app:a_n:asymptotic:3} because we assume that the series 
in \eq{app:a_n:asymptotic:3} is convergent and therefore 
\bea
  \sum_{m\approx n}^{\infty}\, d_m\, \frac{(\nu+m)\,x^m}{\Gamma(1+\nu+m+x)} 
  \label{eq:app:a_n:asymptotic:3:series}
\eea
tends to zero as $n\to\infty$. 

Now we briefly discuss the large $x$ behavior of \eq{app:a_n:asymptotic:3}. 
For this purpose, we first expand $x^m$ in terms of ratios of the gamma functions as follows
\bea
 x^m = \sum_{l=0}^{m} S_{(m,l)} \frac{\Gamma(1+\nu+m+x)}{{\Gamma(1+\nu+l+x)}}\, ,
  \label{eq:app:x^m:expansion}
\eea 
where $S_{(m,0)}=1$, $S_{(m,1)} = -m(\nu+\frac{m+1}{2})$ and so on. 
Plugging \eq{app:x^m:expansion} to \eq{app:a_n:asymptotic:3} 
and changing the order of sums over $m$ and $l$, we find 
\bea
  a_n(x) &\sim& \, \Gamma(\nu+n+x)\,x^{-n}\,\sum_{l=0}^\infty \frac{1}{\Gamma(1+\nu+l+x)}\,
  \sum_{m=l}^{\infty}\,(\nu+m)\, d_m\, S_{(m,l)}  \nonumber \\
         &\sim& \, \Gamma(\nu+n+x)\,x^{-n}\,\left. \left( \frac{-\frac{d}{dz}f_0(z)}{\Gamma(1+\nu+x)}
		+ \cdots \right)\right|_{z=1} 
  \label{eq:app:a_n:asymptotic:4}
\eea 
as  $n\to\infty$.  
This relation can be used to investigate the large $x$ behavior of $a_n(x)$.   

\end{appendix}

\clearpage
\bibliographystyle{apsrev4-1}
%

\end{document}